\documentclass[12pt]{article}
\usepackage{epsfig}
\usepackage[hang,bf,small]{caption}
\usepackage{amsmath}
\DeclareGraphicsExtensions{.eps.gz,.eps,.ps,.ps.gz}
\parskip=\itemsep               

\newcommand{\beq}{\begin{equation}}
\newcommand{\eeq}{\end{equation}}
\newcommand{\be}{\begin{eqnarray}}
\newcommand{\ee}{\end{eqnarray}}

\newcommand{\D}{\partial}

\newcommand{\as}{\alpha_S}
\newcommand{\Lb}{\left(}
\newcommand{\Rb}{\right)}

\newcommand{\half}{\textstyle\frac{1}{2}}
\newcommand{\rv}{\vec{r}} 
 
\newcommand{\hrpar}{\hat{r}_{||}} 
\newcommand{\bv}{\vec{b}} 
\newcommand{\scrbx}[1]{\mbox{\scriptsize #1}}
\def\eq#1{{Eq.~(\ref{#1})}}
\def\fig#1{{Fig.~\ref{#1}}}

\def\gevm{{\,\mbox{GeV}^{-1}}} 
 
\newcommand{\bas}{\bar{\alpha}_S}
\relax
\begin{document}
\title{
{\Large \bf  A modified Balitsky-Kovchegov equation }
}
\author{
{\large  ~E. ~Gotsman, \thanks{e-mail:
gotsman@post.tau.ac.il}~~${}^{a)}$ E.~Levin,\thanks{e-mail:
leving@post.tau.ac.il;\,\,\,levin@mail.desy.de}~~${}^{a),\,b)}$
~ ~U. ~Maor\thanks{e-mail:
maor@post.tau.ac.il}~~${}^{a)}$ ~\,and\,
~ ~E. ~Naftali \thanks{e-mail:
erann@post.tau.ac.il}~~${}^{a)}$
}\\[4.5ex]
{\it ${}^{a)}$ HEP Department}\\
{\it  School of Physics and Astronomy}\\
{\it Raymond and Beverly Sackler Faculty of Exact Science}\\
{\it Tel Aviv University, Tel Aviv, 69978, ISRAEL}\\[2.5ex]
{\it ${}^{b)}$ DESY Theory Group}\\
{\it 22603, Hamburg, GERMANY}\\[4.5ex]
}

\date{\today}
\maketitle
\thispagestyle{empty}

\begin{abstract}
     We propose a modified version of the Balitsky-Kovchegov (B-K) 
evolution equation, which includes the main NLO corrections. We use the 
result that the main NLO corrections to the BFKL kernel are the LO DGLAP 
corrections.
      We present a numerical solution of the modified nonlinear equation, 
and compare with the solution of the unmodified B-K equation. We show 
that the saturation momentum has a sharp increase in the  LHC energy 
range. Our numerical solution shows that the influence of the 
pre-asymptotic corrections, related to the full anomous dimensions of the 
DGLAP equation, are rather large. These corrections moderate the energy 
behaviour of the amplitude, as well as the value of the saturation scale.
All our calculations are made with a fixed value of $\alpha_{s}$. 

 \end{abstract}
\thispagestyle{empty}
\begin{flushright}
\vspace{-20.5cm}
TAUP-2713-02\\
DESY 02-133\\
\today
\end{flushright}
\newpage
\setcounter{page}{1}

\section{Introduction}
\setcounter{equation}{0}

High density QCD  has a long history, starting with Refs.
 \cite{GLR,MUQI,MV}, and is now entering a new phase of its development:
 direct comparison with the experimental data \cite{GW,KLMN}.  This makes
 stringent demands on the theoretical approach to provide reliable
 predictions. The main theoretical tool is the non-linear Balitsky-Kovchegov
 (BK) evolution equation \cite{BK} which has a few deficiencies:
\begin{itemize}
 \item \quad It is correct only in the  leading $\log (1/x)$ 
approximation of
perturbative QCD (pQCD). In practical terms it means that the kernel of 
the  BK
equation should be the BFKL kernel\cite{BFKL} in the leading order (LO) of
pQCD;
 \item \quad This equation is  the mean field approximation to the JIMWLK
equation\cite{JIMWLK}, which is more general, but also more complicated.
  The JIMWLK equation has not been sufficiently well
investigated theoretically, to attempt any description of the 
experimental
data;
 \item \quad We know that the BK equation is not correct in the saturation
region\cite{IM,LL}, hence cannot be viewed as a reliable tool to explore this
region;
 \item \quad The region where we can neglect non-linear corrections, should
be specified by conditions which are beyond the BK equation\cite{MUSH}.
\end{itemize}

In this paper we address only one of the above problems, namely, the higher order
 corrections to the kernel of the BK equation.

As it has been discussed \cite{GLR,MUQI,MV,BK}, the non-linear corrections
 should be incorporated {\em before} taking into account the next-to-leading
 order (NLO) corrections to the kernel of the linear equation (BFKL or DGLAP
 \cite{DGLAP}).  The non-linear corrections are proportional to the exchange
 of two BFKL ladders \cite{GLR,MUQI}. Since the 
 dipole amplitude in the
 BFKL approach has the following behaviour
\beq \label{NLL1}
N_{\scrbx{non-linear term}}\,\,\,\propto\,\,\,\as^4 \,s^{2 \Delta_{BFKL}}\,,
\eeq
where, N denotes the dipole-target amplitude and behaves as  
\beq \label{NLL2}
 N_{\scrbx{linear term}}\,\,\,\propto\,\,\,\as^2 \,s^{ \Delta_{BFKL}}
\eeq
and  $\Delta_{BFKL}$ is the intercept of the BFKL Pomeron ($\Delta_{BFKL}
 \,\,\propto\,\,\as$ in the leading order).

Comparing \eq{NLL1} and \eq{NLL2} one sees that for a wide range of
 energies one has to include the non-linear effects when
\beq \label{NLL3} 
y\,\,=\,\,\ln s\,\,>\,\,\frac{2}{\as}\,\ln \Lb \frac{1}{\as} \Rb 
\eeq

On the other hand, the corrections of the order of $\as^2$ in
 $\Delta_{BFKL}$, becomes essential only for energies $\as^2 y \,\approx\,1$
 or $y\,\,=\,\,\ln s\,\,>\,\,\frac{1}{\as^2}$. Comparing this estimate with
 \eq{NLL3} we see that the correct theoretical strategy is to take
 into account all non-linear corrections first, and then to calculate 
the
 NLO corrections to the BFKL kernel.

However, explicit calculations of the NLO corrections to the BFKL kernel
 \cite{NLL}, show that $\as^2$ corrections to $\Delta_{BFKL}$ are
 rather large, and for the past five years,  have been the subject of
 many debates and discussions \cite{NLLG}.  
The correct resummation of the NLO corrections have
 been performed  by the Florence group
 (see Ref. \cite{FG}),
 and it turns out that the  NLO corrections are 
essential in
 estimates of the non-linear effects. As has been discussed in
 Refs. \cite{GLR,BALE,MUPE,DG} the estimates for the saturation momentum does
 not depend on the precise structure of the non-linear
 corrections. Mueller and Triantafyllopoulos found in Ref.~\cite{MUTR}, 
that the 
NLO
 corrections  crucially change the energy behaviour of the saturation 
scale,
 making it close to the phenomenological saturation scale  appearing 
in
 the Golec-Biernat and Wusthoff model \cite{GW}.

 At first sight, it seems that we are in a vicious circle: on the one 
hand,
 the NLO corrections are important, while on the other hand, they should be
 taken into account only after non-linear corrections. At the moment,
  we have no idea of how to combine these two effects into one
 equation. In this situation  both the non-linear BK equation in the 
LO,
 and the linear BFKL equation in NLO are unreliable.

In this paper we suggest a way out of this dilemma.  We propose a 
modified
 version of the BK equation that takes into account the main NLO corrections.
 The idea on which this equation is based, is the fact that the main
 next-to-leading order corrections to the BFKL kernel are actually the
 leading order DGLAP corrections \cite{FG,DG}.  Therefore, in our
 equation we suggest a procedure in which the non-linear evolution equation
 based on the LO BFKL kernel, is combined with LO DGLAP evolution. 
Different
 approaches of how to make such matching, have been discussed both for 
the
 linear equation (see for example Refs.~\cite{GLR,CCFM} ), and for 
the non-linear
 equation (see Refs.~\cite{GLMF2,DURF2,KS,LU1} and references therein). We 
trust
 that we have found an economical  and  simple method to include 
LO DGLAP
 corrections in the framework of the non-linear equation.  The equation is
  discussed  in  detail in section 2.  Section 3 is devoted to the
 numerical solution of the modified non-linear equation, and to a 
comparison
 with the solution of the BK equation without any modification.  In the
 Conclusion section 4, we summarize the results and discuss the 
application of
 the new equation to a global fit of the experimental data.

\section{The modified non-linear equation}
\setcounter{equation}{0}

\subsection{The saturation scale}

In this section we discuss the value of the saturation scale, which does 
not
 depend on the form of the non-linear term in the BK equation.  The linear
 part of the evolution equation can be written in a simple form, if we go 
to
 the double Mellin transform for the dipole scattering amplitude ($N$), 
namely,
\beq \label{DMELL}
N \Lb \xi \,=\,\ln(r^2/R^2),Y = \ln(1/x); b \Rb \,=\,\,\int \,\frac{d \omega}{2 \pi\,i}\,\frac{d \gamma}{2 
\pi\,i}\,N(\omega,\gamma; b)\,e^{ \omega\,Y\,\,+\,\,\gamma\,\xi}
\eeq
 where $r$ and $R$ are, respectively, the sizes of the dipole and the 
target.  $x$
 denotes the Bjorken variable for the dipole-target scattering, and $b$ 
is the
 impact parameter for the reaction.  The BFKL equation (or in other words the
 linear part of the BK equation) has the form:
\beq \label{BFKLOM}
\omega\,\,\,=\,\,\omega(\gamma)\,\,=\,\,\bas\,\chi_{LO}(\gamma)\,\,+\,\,\bas^2\,\chi_{NLO}(\gamma)
.
\eeq
 For $\chi_{LO}(\gamma)$ we have the well known expression
\cite{BFKL}:

\beq \label{CHILO}
\chi_{LO}(\gamma)\,\,=\,\,2\,\psi(1) \,\,-\,\,\psi(\gamma)\,\,-\,\,\psi(1 - \gamma)\,\,=\,\,\frac{1}{\gamma} + \frac{1}{ 1 - \gamma} 
\,\,+\,\,\chi^{HT}_{LO}(\gamma),
\eeq

with $\psi = d \ln \Gamma (\gamma)/d \gamma$ and $\Gamma$
  the Euler Gamma function. We will discuss the form of 
$\chi_{NLO}(\gamma)$
 later. $\chi^{HT}_{LO}(\gamma) = 2\,\psi(1) \,\,-\,\,\psi(1 +
 \gamma)\,\,-\,\,\psi(2 - \gamma)$ denotes  the contribution of the higher
 twist which we will discuss below.

The form of \eq{BFKLOM} is clear since the  l.h.s. and r.h.s. of the BFKL 
equation
\beq \label{BFKL}
\frac{\partial\, N(Y,\xi)}{\partial\,Y}\,\,=\,\,\int\,K_{BFKL}(\xi,\xi')\,N(Y,\xi')\,d\,\xi
\eeq
have the following Mellin transforms:
\be
\frac{\partial\, N(Y,\xi)}{\partial\,Y}\,&\,\,\rightarrow\,\,&\,\omega N(\omega,\gamma) \label{MELI1}\\
\int\,K_{BFKL}(\xi,\xi')\,N(Y,\xi')\,d\,\xi\,&\,\,\rightarrow\,\,&\,\omega(\gamma)\, 
N(\omega,\gamma)\label{MELI2}.
\ee

The saturation momentum can be calculated without making any assumption 
regarding
 the character of the non-linear corrections, and it has the following 
form:
\be \label{TEQS}
Q^2_s(Y)\,\,&=&\,\,Q^2_s(Y_0)\,\exp  \left(  \,\frac{\omega(\gamma_{cr})}{1 - \gamma_{cr}}\,\,(Y - Y_0)\,\,
- \,\,\frac{3}{2 ( 1 -
\gamma_{cr})}\,\ln(Y/Y_0) -  \right.  \nonumber\\
\, &-&\,\left. \frac{3}{( 1 - \gamma_{cr})^2}\,\sqrt{\frac{2\,\pi}{\omega''(\gamma_{cr})}}\,
( \frac{1}{\sqrt{Y}}\,-\, 
\frac{1}{\sqrt{Y_0}}\,)\,\right),
\ee
where $Y = \ln(1/x)$ is our energy variable, $\omega''(\gamma)= d^2
\omega(\gamma)/(d \gamma)^2$, the value of $\gamma_{cr}$ can be found
from the equation \cite{GLR,MUTR}:
\beq \label{GAMMACR}
\frac{\omega(\gamma_{cr})}{ 1 - \gamma_{cr}} \,\,\,=\,\,\,- \,\,\frac{d \omega(\gamma_{cr}}{ d \gamma_{cr}}\,,
\eeq
 we have normalized the value of the saturation momentum at $Y = Y_0$.

\begin{figure}[ht]
    \begin{center}
        \includegraphics[width=0.70\textwidth]{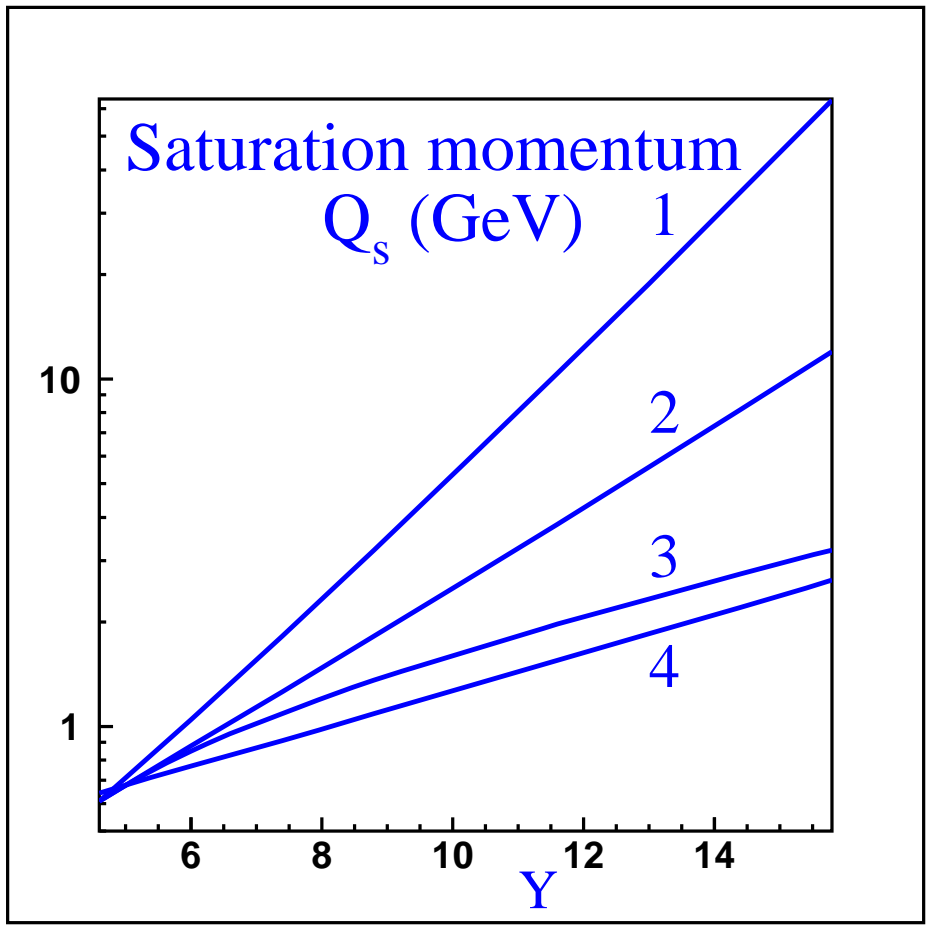}
        \caption{\it The saturation momentum as a function of $x$ ($Y = 
\ln(1/x)$) in the 
LO BFKL (curve 1) ,
 for  \eq{NLOOUR}
(curve 2) and
for \eq{NLOKDG} (curve 3)  for $\bas = 0.2$. Curve 4 is the Golec-Biernat and Wusthoff
 phenomenological saturation momentum. }
\label{qste}
    \end{center}
\end{figure}

The first term was given in the GLR paper \cite{GLR}, the second in Ref.
 \cite{MUTR}, and the third in Ref. \cite{MUPE}.  Fixing the value of
 $Q_s(Y_0) = 0.37 \,GeV^2$ at $x=10^{-2} (Y_0 =4.6) $, we obtain the
 saturation momentum shown in \fig{qste} for $\as = 0.2$ (see curve 1).  
We
 see that the saturation momentum shows a steep rise towards the LHC 
energies. Consequently, prior to
  any discussion of the value and energy dependence of 
the dipole
 scattering amplitude, we need to understand how the next-to-leading order
 corrections could effect the value of the saturation scale. As we have
 mentioned, these corrections change the value of the scale considerably
 \cite{MUTR,DG}.  This can be seen  in \fig{qste}, comparing curve 1 with 
curve
 3 which shows the behaviour of the saturation scale for the NLO BFKL kernel.

\subsection{The NLO BFKL kernel}

 We now discuss the NLO corrections, and suggest a procedure to include
 these in the equation, in a manner so that collective effects are 
treated properly.
   The NLO BFKL kernel is known \cite{NLL}, and all theoretical
 problems related to this kernel have been solved \cite{FG}.

 Our procedure is based upon the important observation, made by the
 Durham group
 \cite{DG}, according to which the main NLO corrections to the BFKL 
kernel,
 can be taken into account using a simple expression for the NLO BFKL
 kernel:
\beq \label{NLOKDG}
\bas\,\chi_{NLO}(\gamma)\,\,\,=\,\,\,\Lb 
\frac{1 + \omega\,A_1(\omega)}{ \gamma} - \frac{1}{\gamma} + \frac{1 + 
\omega\,A_1(\omega)}{1 -  \gamma + \omega} - \frac{1}{ 1 - \gamma}\,\Rb \,\,-\,\,\omega \,
\chi^{HT}_{LO}(\gamma)\,,
\eeq
where,
\beq \label{AOM}
A(\omega) = - 11/12 + O(\omega) + n_F \Lb \frac{\bas}{4\,N^2_c\,\gamma}\,P_{qG}(\omega)\,P_{qG}(\omega) - 1/3 \Rb\,,
\eeq
with $P(\omega)$ being the DGLAP kernel.

The singularities in \eq{NLOKDG} describe the different branches of evolution
 corresponding to 
 the sizes of the interacting dipole (or the  transverse momenta of
 partons). The pole at $\gamma=0$ corresponds to the normal twist-2 DGLAP
 contribution, with the ordering in the transverse parton momenta $Q > 
\dots
 k_{i,t} > k_{t,i+1} > \dots > Q_0$, where $Q_0$ is the typical 
virtuality of
 the target $Q_0 \approx \,1/R$. The pole at $\gamma = 1$ in \eq{NLOKDG}
 corresponds to inverse $k_t$ ordering ($ k_{i,t} < k_{t,i+1} < \dots
 \,Q_0$).  The other poles, at $\gamma = -1, -2, \dots (\gamma = 2,3,
 \dots)$, are the higher twists contributions due to the gluon reggeization\footnote{%
 One of us (E.L.)  acknowledges long and fruitful  discussions
     with J. Bartels            on the
 physical meaning of all parts of the BFKL kernel. }.

As can be seen in \eq{NLOKDG}, the main alterations that we need to make 
to incorporate NLO,  
 are in the sector which corresponds to the ordinary DGLAP evolution. 
We need to change
 the BFKL anomalous dimension, so as to account for the DGLAP anomalous 
dimension.
 The second important change that the NLO calculations induce, is in the
 inverse evolution. However, this branch of evolution is moderated by the
 non-linear effect, rather than by the NLO corrections.  Indeed, it was 
shown
 in Ref. \cite{LT} that the term $1/(1 - \gamma)$ leads to  exponentially
 small corrections in the saturation region, or, in other words, for 
$\gamma >
 \gamma_{cr} \,\approx\,0.37$. Therefore, in attempting to find a new 
kernel for the
 non-linear equation, we can neglect changes in the inverse evolution, and
 keep the BFKL kernel without a shift from $\gamma $ to $\gamma - \omega$.

Using the observation of Ref. \cite{EKL}, according to which \eq{AOM} for
 $A_1(\omega)$ can be well approximated to an   accuracy ($> 95\%$), 
by
 $A_1(\omega) = 1$.  We  rewrite the NLO BFKL kernel in a very 
simple form:
\beq \label{NLOOUR}
\chi_{NLO}(\gamma) \,\,\,=\,\,\, - \omega\,\chi_{LO}(\gamma)\,.
\eeq
 It should be mentioned that $A_1(\omega) = 1$ corresponds to the
 expression for the anomalous dimension of the DGLAP equation:
 \beq \label{ADEKL}
 \gamma_{LO}(\omega)\,\,=\,\,\bas\,\left(\,\frac{1}{\omega}\,\,-\,\,1\,\right)\,,
 \eeq 
 which approximates the DGLAP anomalous dimension in  leading order to  
within
 an  accuracy of 95\%. The actual deviation of \eq{ADEKL} from the 
correct expression 
for the DGLAP anomalous dimension in  leading order, is even less than 
5\%.

The saturation momentum for \eq{NLOOUR} is plotted in \fig{qste} (see
 curve~2).  This momentum is larger than the saturation
 momentum given by \eq{NLOKDG}, but we trust that the non-linear 
corrections
 will suppress it.

\subsection{Modified non-linear equation} 

Using \eq{NLOOUR} we obtain the full kernel for the linear equation in the
 form: 
\beq \label{KERF} \omega(\gamma) \,\,=\,\,\bas \,\chi_{LO}(\gamma)\,\Lb
 1 \,\,-\,\,\omega \Rb 
\eeq

This kernel imposes  energy conservation (see Ref. \cite{EKL}), and
 describes the NLO BFKL kernel. It does not include  the contribution 
coming from 
 inverse ordering, which should be suppressed in the solution to the
 non-linear equation.

 Our suggestion is to use \eq{KERF} as the kernel for the
 non-linear equation. 
 
The modified non-linear equation with this kernel has the form
$$
\frac{\partial N\Lb r,Y;b \Rb}{\partial\,Y}\,+\,\ \frac{C_F\,\as}{\pi^2}\,\,\int\,\frac{d^2 r'\,r^2}{(\vec{r} 
\,-\,\vec{r}\:')^2\,r'^2} \,\frac{\partial N\Lb r',Y;\vec{b} - \frac{1}{2}\,(\vec{r} - \vec{r}\:')\Rb}{\partial\,Y}
\,\,
=\,\,\frac{C_F\,\as}{\pi^2}\,\,\int\,\frac{d^2 r'\,r^2}{(\vec{r}\,-\,\vec{r}\:')^2\,r'^2}\,
$$
\beq \label{MODBK}
\Lb \,2  N\Lb r',Y;\vec{b} \, -\ 
\frac{1}{2}\,(\vec{r} - \vec{r}\:')\Rb\,\,-\,\, N\Lb r',Y;\vec{b} - \frac{1}{2}\,(\vec{r} - \vec{r}\:') \Rb\, N\Lb
 \vec{r} - 
\vec{r}\:',Y;b - \frac{1}{2} \vec{r}\:'\Rb  \,
\Rb.
\eeq

  Using \eq{MELI1} and \eq{MELI2} one can recognize , 
that the r.h.s. of 
the equation is the usual BK 
equation,
 while the second term on the l.h.s. of the equation is a new 
contribution
 which takes into account our modification of the BFKL kernel. Indeed, 
the  
term
 $\bas\,\,\omega 
\,\,\chi_{LO}(\gamma)$ has the following form in  the $Y,r$ 
representation
\beq \label{MELI3}
\bas\,\omega\,\,\chi_{LO}\Lb\gamma \Rb \,\,\,\,\rightarrow\,\,\,\,\bas\,\int 
\,K_{LO}\Lb Y, r' \Rb 
\,d^2\,r'\,\frac{\partial N\Lb Y,r' \Rb}{\partial\,Y}.
\eeq

In \eq{MODBK} the linear part conserves energy, however, in the 
non-linear part
 energy is not conserved. It should be stressed that although
 \eq{MODBK} is written in
 the leading order of $\as$,  it also includes  terms that 
are not
 proportional to $ (\as \ln(1/x))^n$.  Such resummation is legitimate, and
 this type of correction could be taken into account before considering the
 non-linear corrections.

In the dipole approach the kernel of the non-linear equation is the 
  probability for one dipole (with size $r$) to decay into two dipoles
 (with sizes $r'$ and $|\vec{r} - \vec{r}\:'|$).  Assuming that this physical
 interpretation is correct in the NLO, we can rewrite \eq{MODBK} in a 
more general
 form:

\beq \label{MODBK1}
\frac{\partial N\Lb r,Y;b \Rb}{\partial\,Y}\,\,\,=\,\,
 \frac{C_F\,\as}{\pi^2}\,\,\int\,\frac{d^2 r'\,r^2}{(\vec{r}
\,-\,\vec{r}\:')^2\,r'^2} \,
\eeq
$$
\Lb \,2  N\Lb r',Y;\vec{b} \, -\
\frac{1}{2}\,(\vec{r} - \vec{r}\:')\Rb\,\,-\,\, N\Lb r',Y;\vec{b} - \frac{1}{2}\,(\vec{r} - \vec{r}\:') 
\Rb\, N\Lb \vec{r} -
\vec{r}\:',Y;b - \frac{1}{2} \vec{r}\:'\Rb  \,
\Rb\,\,-
$$
$$
\,-\,\frac{\partial}{\partial\,Y} \Lb \,2  N\Lb r',Y;\vec{b} \, -\
\frac{1}{2}\,(\vec{r} - \vec{r}\:')\Rb\,\,-\,\, N\Lb r',Y;\vec{b} - \frac{1}{2}\,(\vec{r} - \vec{r}\:') 
\Rb\, N\Lb \vec{r} -
\vec{r}\:',Y;b - \frac{1}{2} \vec{r}\:'\Rb  \Rb.
$$
 The advantage of this equation is that energy is conserved by the non-linear
 term, as well as by the linear one.

\subsection{Pre-asymptotic corrections to the value of the saturation scale}

Before solving \eq{MODBK1} numerically, we would like to justify the need 
for
 a numerical solution.  Indeed, at first sight we have \eq{TEQS} (see also
 \eq{QSRA} below) for the saturation scale, and we have a good understanding
 of (i) how a solution to BK equation approaches the saturation boundary $N =
 1$ (see for example Ref. \cite{LT} ), and (ii) how the amplitude behaves in
 the vicinity of the saturation scale \cite{MCLIGS,MUTR}.

We believe that the main shortcoming of our analytical approach at present,
 is the fact that it can only be applied for asymptotically high 
energies,
 while for any practical use we need to know the solution at rather low
 energies.  To specify what high and  low are (for this particular
 problem), we return to the discussion of the value of the saturation 
scale
 given by \eq{TEQS}.  However, we  now consider the equation from a 
different
 point-of-view, and ask ourselves, how large are the low energy corrections
 to the value of the saturation scale. 

\begin{figure}[ht]
    \begin{center}
       \includegraphics[width=0.60\textwidth]{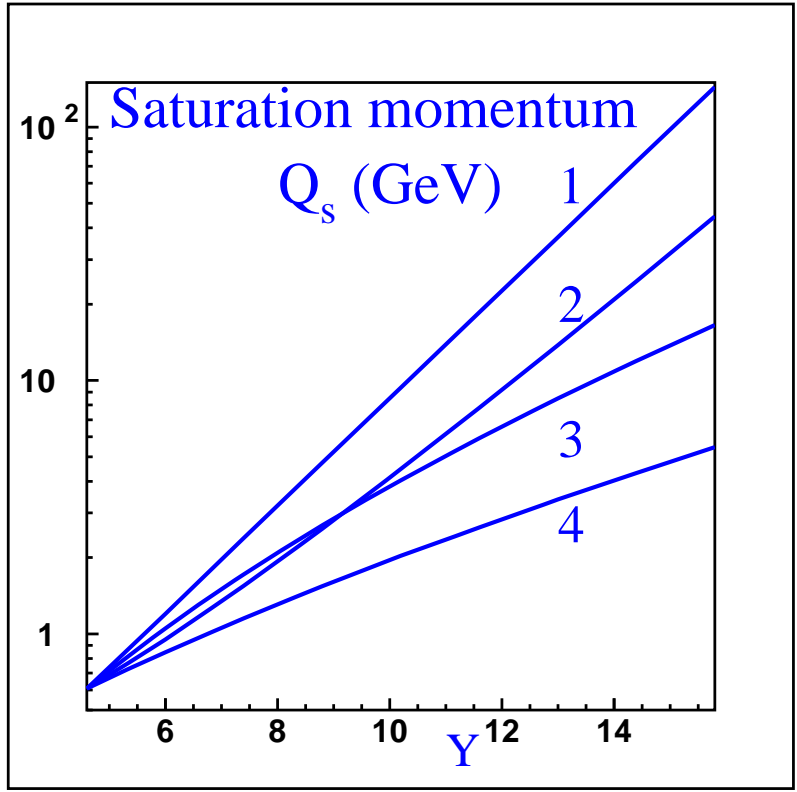}
        \caption{\it The saturation momentum as function of $x$ ($Y = 
\ln(1/x)$ ). 
Saturation momenta for the case of fixed  QCD coupling ($\as = 0.2$): 
curve 1 shows the high energy 
behaviour of $Q_s$ (the first term in \eq{TEQS} ), while curve 2 shows all 
term in  \eq{TEQS}. Curves 3 
and 4 are the same as curve 1 and 2 but for running $\as$. Curve 3 shows 
the first term of \eq{QSRA} which describes the high energy
 behaviour of the saturation scale
curve 4 is calculated using all terms of \eq{QSRA}.}
 \label{qscom}
    \end{center}
\end{figure}

\fig{qscom}, shows the saturation scale when only the high energy
 contribution (the first term in \eq{TEQS}) is taken into account (curve 1),
 and the saturation scale including the low and high energy corrections
 (curve 2). One can see that the difference is very large, even for the 
LHC
 energy range.  The same situation  holds  for the case of 
the
 running $\as$.  In this case the formula for the saturation scale with the
 low energy corrections is:
\beq \label{QSRA}
Q^2_s \left( Y \right)\,\,= 
\eeq
$$
\,\,Q^2_s \left( Y_0 \right)\,\,\exp \left( \sqrt{\frac{8\,N_c 
\,\chi(\gamma_{cr})}{b\,\gamma_{cr}}}\,(\,\sqrt{Y}\,-\,\sqrt{Y_0} 
)\,\,-\,\,2.338\,\frac{3}{4}\,
\left(\frac{\chi''(\gamma_{cr})
    \sqrt{2\,N_c}}{\sqrt{b\,\gamma_{cr}\,\chi(\gamma_{cr})}}\right)^{\frac{1}{3}}\,\,
(\,Y^{\frac{1}{6}}  
\,-\,Y^{\frac{1}{6}}_0 ) 
\right).
$$

 The energy behaviour of the first term of \eq{QSRA} (high energy), and 
for both
 terms are shown in curves 3 and 4 of \fig{qscom}. The difference between
 these two curves, which is large, indicates the size of the low energy
 correction.

The same conclusion regarding the essential low energy corrections to the 
value
 of the saturation, and its energy behaviour, has been derived from the
 numerical solution of the BK equation.  In Ref. \cite{LUB} the energy
 dependence of the saturation scale in the LHC range of energies turns out to
 be more moderate, than predicted by asymptotic formulae of the type of
 \eq{TEQS}.

 We can therefore conclude, that to provide reliable predictions for the
 LHC range of energies, it is necessary
 to solve the BK modified equation
  numerically  at low energies. 

The second conclusion, which we can derive from \fig{qscom}, comparing curve
 4 and curve 2, is the fact that the inclusion of the running QCD coupling is
 essential, so as to provide a reliable extrapolation of the HERA data to
 the LHC energy range.

\section{Numerical Solution}
\setcounter{equation}{0}

Ideally, our primary goal would have been to obtain a numerical solution to
 the full, energy conserving, modified BK equation.   Although,
 \eq{MODBK1} is a well defined integro-differential mathematical equation,
 obtaining a numerical solution to this equation is a rather complicated 
task,
  due to the appearance of $\D N/\D Y$ on both sides of the equation.

A natural first step towards a numerical solution of \eq{MODBK1} is to
 rewrite the DGLAP correction in the double log approximation (the second term in l.h.s. of \eq{MODBK}), 
including only
 leading twists, and neglecting the impact parameter dependence.  We 
follow
 Ref.~\cite{GLMSOLB}, and modify the kernel of the BK terms in the 
modified
 equation so as to exclude contributions of inverse fan diagrams.

We  write the evolution equation in a differential form as:
\begin{eqnarray} \label{fixb}
\lefteqn{
\frac{\partial N\Lb r,Y;b \Rb}{\partial\,Y} =
\frac{C_F\,\as}{\pi^2}\,\,
\left\{\rule{0ex}{4ex}
-\,2r^2\,\int_{r}^R\,\frac{d^2 r'}{r'^4} \,
\frac{\partial N\Lb r',Y;b\Rb}{\partial\,Y}\,+\right.}& &
\nonumber \\ 
& &
\int\,\frac{d^2
  r'\,r^2}{(\vec{r}\,-\,\vec{r'})^2\,r'^2}\,\Theta(R-r')\Theta(R-|r-r'|)
\times \nonumber\\
& & 
\left.
\left[ 2  N\Lb r',Y;\vec{b} - \half(\vec{r} - \vec{r'})\Rb - 
          N\Lb r',Y;\vec{b} - \half(\vec{r} - \vec{r'})\Rb\,
          N\Lb \vec{r} - \vec{r'},Y;b - \half\vec{r'}\Rb
\right]\rule{0ex}{4ex}\right\}\,.
\end{eqnarray}
 In our numerical calculations we have taken the value of the target 
size, $R$,
 to $\sqrt{10}\,\gevm$.  Note that, in principle, in the double log
 approximation, DGLAP contributions should be calculated for large distances,
 irrespective of the size of the target.  However, as our pure BK solution
 can only be trusted  for $r < R$ (see \cite{GLMSOLB} for further 
details), we
 impose the same restriction  on the $r'$ integration of the first term
 of the r.h.s of \eq{fixb}.

Our solution to \eq{fixb} will be presented in terms the integrated
quantity:
\begin{equation}\label{eq:int-theta}
N(Y,r;b) \equiv\- \int d\hrpar\phantom{-} N(Y,r;b;\hrpar)\,, 
\end{equation}
 where $r\equiv|\rv|$, $b\equiv|\bv|$, and $\hrpar\equiv\bv\cdot\rv/(b r)$.
 This quantity determines the  physical observables in DIS, such as $F_2$ 
and
 gluon densities.

Briefly, our numerical procedure is as follows.  Denoting  a
 particular rapidity by $Y_i$
, at which the solution and its derivative are known, we
 define the solution at $Y_{i+1}\equiv Y_i+h(Y)$ as a matrix, in which the
 matrix elements correspond to (fixed $Y$) solutions at different $r$ and
 $b$, integrated over $\hrpar$:
\begin{equation}\label{eq:solution1}
N(Y_{i+1},r;b) = N(Y_{i},r;b)+h(Y_i)\int d\hrpar 
\left. \frac{\partial N(Yr,;b;\hrpar)}{\partial\,Y}\right|_{Y=Y_i}\,,
\end{equation}
where $\partial N(Y,r;b;\hrpar)/\partial\,Y$ is given by the r.h.s. of
(\ref{fixb}), and $h(y_i)$ is a variable step size in the rapidity
space.

Our input to \eq{eq:solution1} at $Y=Y_0\approx 4.6$ was taken from the 
fixed
 $b$ solution \cite{GLMF2}, from which we evolved over a range of about 10
 units of rapidity.  The selection of the rapidity difference between two
 successive steps of the evolution was based on the Euler two-step procedure,
 in which, for each $i$, a first solution is obtained along the path
 $Y_i\rightarrow Y_i+h$, and a second solution is obtained along the path
 $Y_i\rightarrow Y_i+h/2\rightarrow Y_i+h$.  The difference per unit step
 between the two solutions is proportional to $h(Y)\D^2N/\D Y^2$.  For each
 step of the evolution, $h(y_i)$ was iteratively reduced until the maximal
 difference between the first and the second solutions over the $i$th matrix
 is small (we set our accuracy condition to $10^{-2}$, and satisfied
 ourselves that the solution is stable within a few percent under variations
 of this choice).

The existence of a first derivative of $N$ on both side of \eq{fixb}, 
slows
 the convergence rate of the equation.  Using parallel programming techniques
 (16 CPUs) and $50\times 50$ matrices, an evolution over 10 units of rapidity
 was completed within about $50-500$ hours, depending on the value of $\as$
 (faster convergence for smaller couplings).  Due to the numerical 
complexity
 of \eq{fixb}, the numerical solution presented below is for fixed $\as$.  We
 recognize, however, that QCD evolution should be performed with running QCD
 coupling.

\begin{figure}
    \begin{center}
        \includegraphics[width=0.9\textwidth, bb=50 180 370 302]{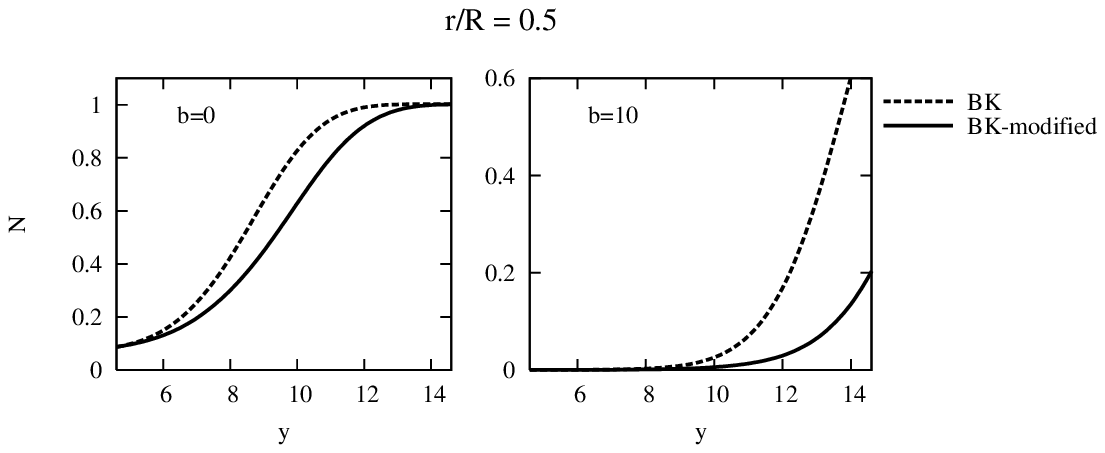}
        \caption{\it 
The $Y$-dependence of the solution to \eq{fixb} for $r/R=\half$ and $b=0,10$
in units of GeV$^{-1}$}
 \label{fig:ydep}
    \end{center}
\end{figure}

\fig{fig:ydep} shows the difference between the solution to the pure and
 modified BK equations.  Shown in \fig{fig:ydep} are the two solutions, as a
 function of $Y,$ at $r/R=\half$ and two values of the impact parameter:
 $b=0$ (left) and $b=10\,\gevm$ (right). The additional term in \eq{fixb}
  moderates the $Y$-dependence of the solution considerably.  The effect 
is
 particularly pronounced for large $b=10\,\gevm$.

\begin{figure}
    \begin{center}
        \includegraphics[width=0.9\textwidth, bb=50 180 370 302]{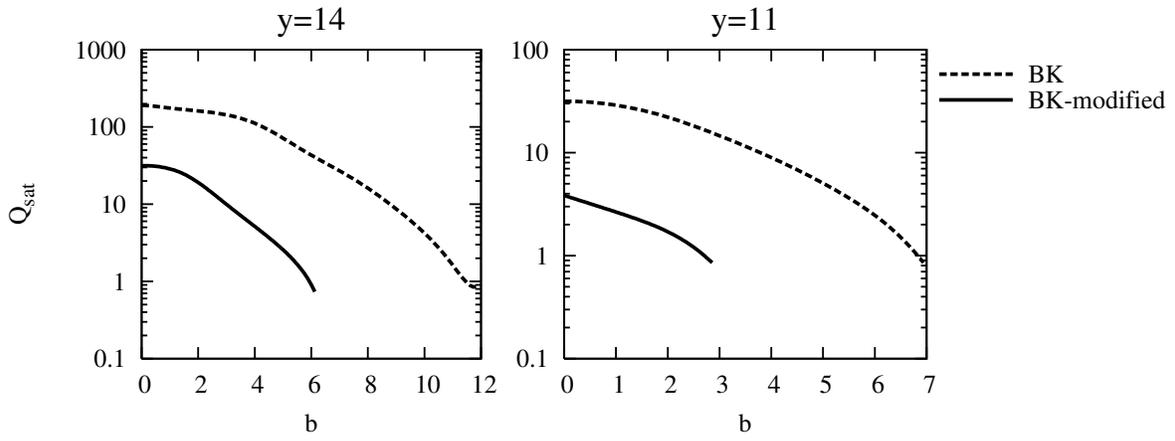}
        \caption{\it 
The $b$-dependence of the saturation scale, b in units of GeV$^{-1}$.
}
 \label{fig:qsat-b}
    \end{center}
\end{figure}

\begin{figure}
    \begin{center}
        \includegraphics[width=0.9\textwidth, bb=50 180 370 302]{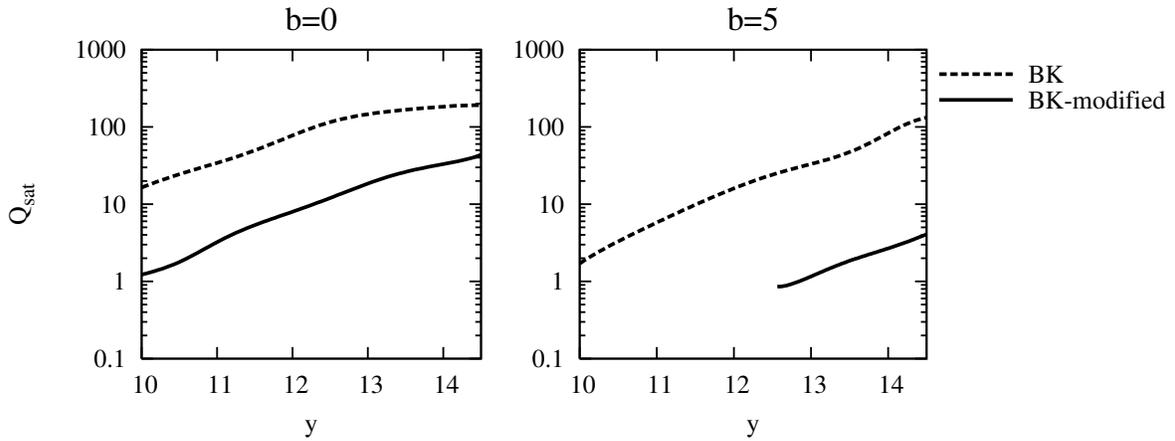}
        \caption{\it 
The $Y$-dependence of the saturation scale.
}
 \label{fig:qsat-y}
    \end{center}
\end{figure}

\begin{figure}
 \begin{center}
        \includegraphics[width=0.9\textwidth, bb=50 50 400 300]{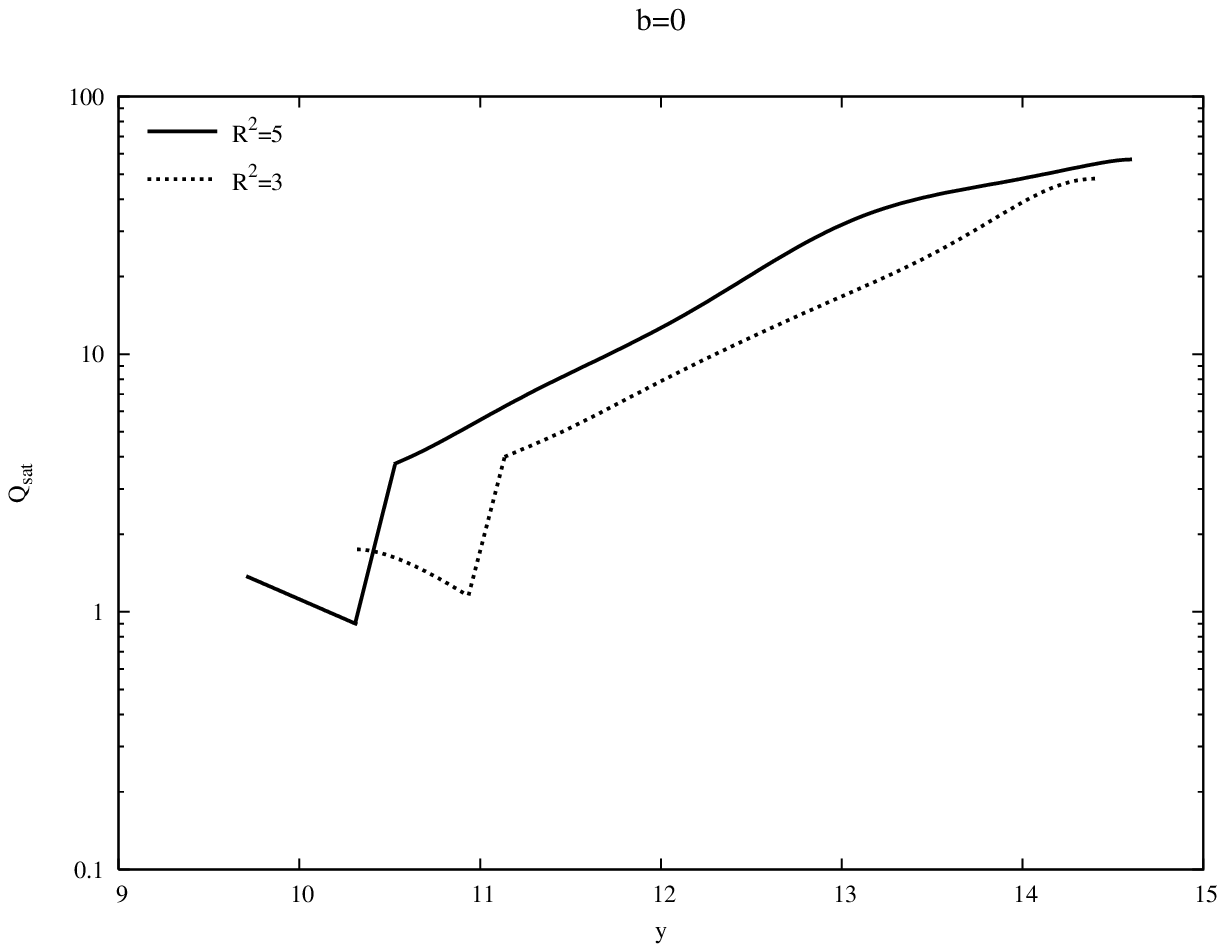}
        \caption{\it 
The effect of the target size, $R$, (in units of GeV$^{-1}$) on the 
solution of the modified BK
equation.
}
 \label{fig:qsat-y2}
    \end{center}
\end{figure}

Figs. \ref{fig:qsat-b} and \ref{fig:qsat-y} show the saturation scale,
 $Q_{sat}(Y,b)$, as a function of $b$ and $Y$, respectively.  The definition
 of $Q_{sat}$ is a matter of taste, and we choose to define it as the value
 of $Q=2/r$ at which $2\log(1-N(Y,r;b))=-1$. We note that both the value and
 the slope of $Q_{sat}$ are significantly reduced with the inclusion of the
 DGLAP term. We have also analyzed the effect of the target size on the
 saturation scale, see \fig{fig:qsat-y2}, where we show $Q_{sat}(Y,b)$ for
 $R^2=3,\,5$ GeV$^{-2}$.  At a certain value of the rapidity, which 
depends on the
 particular definition of the saturation scale and the target size, $Q_{sat}$
 exhibits an abrupt transition from a flat, or even decreasing with 
 energy, to a steep increasing energy dependence.  Such transition was 
also observed in
 \cite{IKMC}, in which the kernel of the BK equation was modified with
 $\exp{(-\lambda |r|)}$, where $\lambda$ is a free parameter.  Broadly
 speaking,  apart from the DGLAP terms and the angle integration, see
 \eq{eq:int-theta}, this approach is similar to a more simplified 
approach we
 adopt here with $\lambda\sim1/R$.

\begin{figure}[t]
\begin{center}
        \includegraphics[width=0.9\textwidth, bb=50 180 370 302]{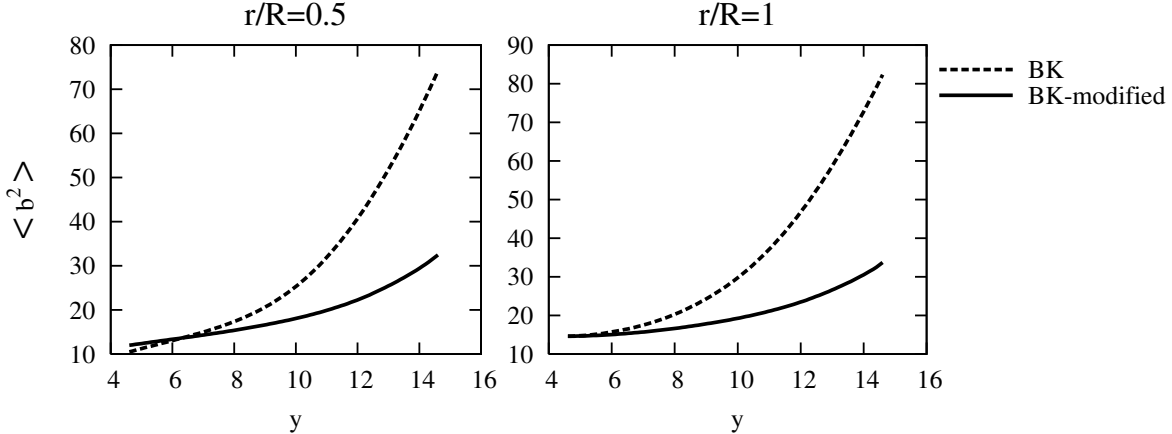}
        \caption{\it 
 Energy behavior of the average $\langle b^{2}\rangle$, see \eq{eq:bav}.
}
 \label{fig:bav}
    \end{center}
\end{figure}

Perhaps, the most reliable test of the validity of the non-linear
 evolution approach would be a comparison to the $F_2$ data.  However,
 the energy dependence of the solutions \cite{GLMSOLB,GBSSOL} to the
 pure BK equation are far too steep to allow such a comparison.  The
 calculation of $F_2$ involves integrating  the dipole amplitude
 over the impact parameter. The increase of $F_2$ with  rapidity can
 therefore be assessed by calculating the $Y$-dependence of $\langle
 b^2\rangle$, defined as:
\begin{equation}\label{eq:bav}
\langle b^2\rangle = \frac{\int d^2b b^2 N(Y,r;b)}{\int d^2b N(Y,r;b)}.
\end{equation}
\fig{fig:bav} shows $\langle b^2\rangle$ as a function of $Y$ for the
solutions to the modified and pure BK equations.  For large $Y=14$ the
overall effect of the DGLAP term reaches about 60\%.

\begin{figure}[t]
 \begin{center}
        \includegraphics[width=0.9\textwidth, bb=50 170 370 300]{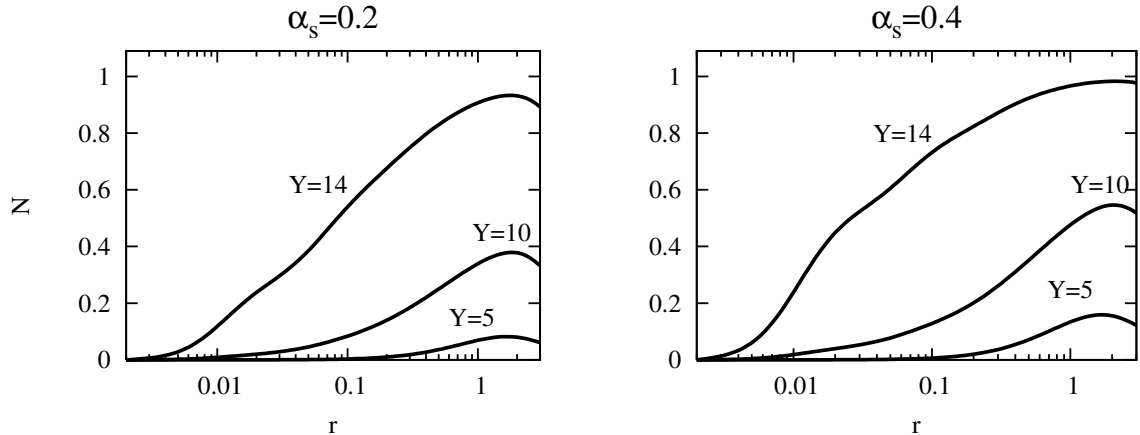}
        \caption{\it 
The evolution process at $b=0$ according to the on the modified BK
equation, for two fixed values of the QCD coupling.
}
 \label{fig:rdep}
    \end{center}
\end{figure}

As stated, QCD evolution should be performed, in principle, for running
 $\as$. Such a numerical procedure  is beyond the scope
 of the present 
 paper.  Nevertheless, one can assess the influence of the 
QCD
 coupling on the evolution process, by performing the evolution with
 different (fixed) values of $\as$.  \fig{fig:rdep} shows the evolution
 process at $b=0$, for $\as=0.2$ and $\as=0.4$.  Note that both $Y$- and
 $r$-dependences are considerably steeper for large QCD coupling.  It is
 much harder to achieve numerical stability for large $\as$.  
 This is due to the DGLAP term in \eq{fixb}, which is proportional to
 $\as\partial N/\partial Y$.  For small $\as$, the DGLAP term is small
 compared to the l.h.s. of the equation, and the error due to the
 discretization (\ref{eq:solution1}) is small.  For larger values of 
$\as$,
 this error becomes greater, resulting in a very small step size 
$h(Y_i)$.  
The
 practical consequence of this is  an increase of about an order of 
magnitude in computing
 time for $\as=0.4$, compared to $\as=0.2$.

\section{Conclusions}
\setcounter{equation}{0}

Our numerical solution to \eq{MODBK}, presented in
 Figs.~\ref{fig:ydep}-\ref{fig:rdep} shows that the influence of the
 pre-asymptotic corrections, related to the full anomalous dimension of the
 DGLAP equation, is rather large. These corrections moderate the energy
 behaviour of the amplitude, as well as the value of the saturation 
scale.
 The BK equation  without any modification does not provide a
 reliable prediction for the LHC energies, while there is 
a reasonable chance that the modified equation 
 will be more successful.

 We note, however, that we have not addressed two issues: (i) the 
solution of
 the full equation \eq{MODBK1}, and (ii) the inclusion of a running QCD
 coupling in the numerical procedure.  In particular, our experience shows
 that it will be necessary to employ
 new numerical methods   to solve 
 \eq{MODBK1}, without which, one cannot find a reliable solution. These
 issues will  be the subject of a future publication.

The running QCD coupling should be used to provide  reliable
 estimates of physical observables in the LHC range of energies (see, for
 example, Refs.~\cite{GLMF2,GLMSOLB} and especially Ref.~\cite{ALAR} where
 the running $\as$ case is discussed in detail). \fig{fig:rdep} indicates
 that running $\as$ will lead to further suppression  of the steep
increasing  energy 
behaviour (see Ref.\cite{ALAR}
 for numerical estimates of this suppression).

We firmly believe that the modified BK equation, proposed in this paper, can
 be a basis for a new global fit of the experimental data, that will 
provide
 reliable predictions for the deep inelastic parton densities, in the LHC
 energy range.

\section*{Acknowledgments}

One of us (EN) would like to thank Ido Adam and Michael Kroyter for fruitful
 discussions.  This research was supported in part by the GIF grant \#
 I-620-22.14/1999, and by the Israel Science Foundation, founded by the
 Israeli Academy of Science and Humanities.



\begin{thebibliography} {99}
\bibitem{GLR}
L. V. Gribov, E. M. Levin and M. G. Ryskin, {\it Phys. Rep.}\, {\bf 100}
(1983) 1.
\bibitem{MUQI}
~A. ~H. ~Mueller and ~J. ~Qiu, {\it Nucl. Phys.} {\bf B268} (1986) 427.

\bibitem{MV}
L. McLerran and R. Venugopalan, {\it   Phys. Rev.}  {\bf D 49} (1994)
2233, 3352; {\bf D 50} (1994) 2225, {\bf D 53} (1996) 458, {\bf D 59}
(1999) 09400.

\bibitem{GW}
K. Golec-Biernat and M. W\"usthoff,
{\it Phys. Rev.} {\bf D59} (1999),
014017; {\it ibid.} {\bf D60} (1999), 114023;
{\it Eur. Phys. J.} {\bf C20} (2001) 313.
\bibitem{KLMN}
D.~Kharzeev, E.~Levin and L.~McLerran,
{\it Phys.\ Lett.}\, {\bf B561} (2003) 93
[arXiv:hep-ph/0210332];\,\,\,
 D. Kharzeev and M. Nardi, {\it Phys. Lett. } {\bf B507},
121 (2001);\,\,\, D. Kharzeev, E. Levin, {\it Phys. Lett.} {\bf B523}, 79
(2001);\,\,\,D.~Kharzeev, E.~Levin and M.~Nardi,
Nucl.\ Phys.\  {\bf A730} (2004) 448
[arXiv:hep-ph/0212316];\,\,\,
 D. Kharzeev, Y. Kovchegov and K. Tuchin,
{\it Phys.Rev.} {\bf D68}, 094013 (2003);\,\,
{\it``Nuclear modification factor in d + Au collisions: Onset of 
suppression 
in the
color glass condensate,''}
arXiv:hep-ph/0405045.

\bibitem{BK}
Ia. Balitsky, {\it Nucl.Phys. } {\bf B463}  (1996) 99;\,\,\,
Yu. Kovchegov, {\it Phys. Rev.} {\bf D60} (2000) 034008.

\bibitem{BFKL}
 E. A. Kuraev, L. N. Lipatov, and F. S. Fadin,{it  Sov. Phys. JETP}
                {\bf 45} (1977) 199 ; \\
Ya. Ya. Balitsky and L. N. Lipatov,
               {\it  Sov. J. Nucl. Phys.}\, {\bf 28} (1978) 22 .
\bibitem{JIMWLK}
~J.~Jalilian-Marian, A.~Kovner, A.~Leonidov and H.~Weigert,
{\it Phys.\ Rev.}\,  {\bf D59} (1999) 014014
[arXiv:hep-ph/9706377];\,\,{\it  Nucl.\ Phys.}\,{\bf B504} (1997) 415
[arXiv:hep-ph/9701284];\\
E.~Iancu, A.~Leonidov and L.~D.~McLerran,
{\it Phys.\ Lett.}\,  {\bf B510} (2001) 133
[arXiv:hep-ph/0102009];\,\, {\it Nucl.\ Phys.}\,  {\bf A692} (2001) 583
[arXiv:hep-ph/0011241];\\
H.~Weigert,
{\it Nucl.\ Phys.}\,  {\bf A703} (2002) 823
[arXiv:hep-ph/0004044].

\bibitem{IM}
E.~Iancu and A.~H.~Mueller,
{\it Nucl.\ Phys.}\,  {\bf A730} (2004) 460, 494,
[arXiv:hep-ph/0308315],[arXiv:hep-ph/0309276].

\bibitem{LL}
E.~Levin and M.~Lublinsky, {\it
Nucl.\ Phys.}\,  {\bf A730} (2004) 191
[arXiv:hep-ph/0308279].




\bibitem{MUSH}
A.~H.~Mueller and A.~I.~Shoshi,
{\it Nucl.\ Phys.}\, {\bf B692} (2004) 175
[arXiv:hep-ph/0402193].

\bibitem{DGLAP}
 V. N. Gribov and L. N. Lipatov, {\it Sov. J. Nucl. Phys} {\bf 15} (1972)
                438;\\
 G. Altarelli and G. Parisi, {\it Nucl. Phys.} {\bf B 126} (1977) 298; \\
Yu. l. Dokshitser, {\it Sov. Phys. JETP} {\bf 46}  (1977) 641.


\bibitem{NLL} V.S.~Fadin and L.N.~Lipatov, {\it Phys. Lett.} {\bf B429}
(1998)
127\\
G.~Camici and M.~Ciafaloni,{\it  Phys. Lett.} {\bf B430} (1998) 349.


\bibitem{NLLG}
 M.~Ciafaloni, D.~Colferai, G.P.~Salam and A.M.~Stasto,
Phys. Rev. {\bf D68} (2003) 114003;\,\,\,
M.~Ciafaloni, D.~Colferai, G.~P.~Salam and A.~M.~Stasto,
{\it
Phys.\ Lett.}\, {\bf B541}, 314 (2002)
[arXiv:hep-ph/0204287];\,\,\,
{\it Phys.\ Rev.}\, {\bf D66}, 054014 (2002)
[arXiv:hep-ph/0204282];\,\,\,
J.~R.~Forshaw, D.~A.~Ross and A.~Sabio-Vera, {\it
Phys.\ Lett.}\, {\bf B498}, 149 (2001)
[arXiv:hep-ph/0011047];\,\,\,
M.~Ciafaloni, M.~Taiuti and A.~H.~Mueller, {\it
Nucl.\ Phys.}\, {\bf B616}, 349 (2001)
[arXiv:hep-ph/0107009];\,\,\,
S.J.~Brodsky, V.S.~Fadin, V.T.~Kim, L.N.~Lipatov and
G.B.~Pivovarov, JETP Lett. {\bf 70} (1999) 155;
D.~A.~Ross, {\it
Phys.\ Lett.}\,\, {\bf 431}, 161 (1998)
[arXiv:hep-ph/9804332];\,\,\,
E.~Levin, {\it Nucl.\ Phys.}\,\, {\bf B545} (1999) 481.
[arXiv:hep-ph/9806228];\,\,\,
N.~Armesto, J.~Bartels and M.~A.~Braun, {\it   
Phys.\ Lett.}\, {\bf 442}, 459 (1998) 
[arXiv:hep-ph/9808340];\,\,\,
Y.~V.~Kovchegov and A.~H.~Mueller,
{\it Phys.\ Lett.}\,\, {\bf B439} (1998) 428   
[arXiv:hep-ph/9805208].

\bibitem{BALE}
J.~Bartels and E.~Levin,
{\it Nucl.\ Phys.}\,  {\bf B387} (1992) 617.
\bibitem{MUPE}
S.~Munier and R.~Peschanski,
{\it ``Universality and tree structure of high energy QCD,''}
arXiv:hep-ph/0401215;\,\,{\it Phys.\ Rev.}\,  {\bf D69} (2004) 034008
[arXiv:hep-ph/0310357];\,\,
{\it Phys.\ Rev.\ Lett.}\,  {\bf 91} (2003) 232001
[arXiv:hep-ph/0309177].

\bibitem{DG}
V.~A.~Khoze, A.~D.~Martin, M.~G.~Ryskin and W.~J.~Stirling,
{\it ``The spread of the gluon k(t)-distribution and the determination of
the
saturation scale at hadron colliders in resummed NLL BFKL,''}
arXiv:hep-ph/0406135.
\bibitem{MCLIGS}
E.~Iancu, K.~Itakura and L.~McLerran,
{\it
Nucl.\ Phys.}\, {\bf A721} (2003) 293; \, {\bf A708}, 327
(2002)
[arXiv:hep-ph/0203137].


\bibitem{MUTR}
A.~H.~Mueller and D.~N.~Triantafyllopoulos,
{\it Nucl.\ Phys.} \, {\bf B640} (2002) 331
[arXiv:hep-ph/0205167];\,\,D.~N.~Triantafyllopoulos,
{\it Nucl.\ Phys.}\,  {\bf B648} (2003) 293
[arXiv:hep-ph/0209121].
\bibitem{FG}
G.P.~Salam, {\it JHEP}\,\, {\bf 9807} (1998) 019;\,\,Acta Phys. Pol. {\bf
B30} (1999) 3679;\\
M.~Ciafaloni, D.~Colferai and G.P.~Salam, {\it Phys. Lett.}\,\, {\bf B452}
(1999) 372; {\it Phys. Rev.}\,\, {\bf D60} (1999) 114036.



\bibitem{CCFM}
M.~Ciafaloni,
{\it Nucl.\ Phys.} \, {\bf B296} (1988) 49;\,\,\,
S.~Catani, F.~Fiorani and G.~Marchesini,
{\it Phys.\ Lett.} \, {\bf B234} (1990) 339;\,\,
Nucl.\ Phys.\ B {\bf 336} (1990) 18.
 \bibitem{GLMF2}
E.~Gotsman, E.~Levin, M.~Lublinsky and U.~Maor,
Eur.\ Phys.\ J.\ C {\bf 27} (2003) 411
[arXiv:hep-ph/0209074], parameterizations are available at www.desy.de/~lublinm/;
M.~Lublinsky, E.~Gotsman, E.~Levin and U.~Maor,
Nucl.\ Phys.\  {\bf A696} (2001) 851
[arXiv:hep-ph/0102321].


\bibitem{DURF2}
M.~A.~Kimber, J.~Kwiecinski and A.~D.~Martin,
{\it Phys.\ Lett.}\, {\bf B508} (2001) 58
[arXiv:hep-ph/0101099].


\bibitem{KS}
K.~Kutak and A.~M.~Stasto,
{\it ``Unintegrated gluon distribution from modified BK equation,''}
arXiv:hep-ph/0408117.
\bibitem{LU1}
G.~Chachamis, M.~Lublinsky and A.~Sabio Vera,
{\it ``Higher order effects in non linear evolution from a veto in rapidities,''}
arXiv:hep-ph/0408333.


\bibitem{LT}  
~E.~Levin and K.~Tuchin, {\it Nucl.\ Phys.}\, {\bf A693} (2001) 787,
[arXiv:hep-ph/0101275]\,;\,\,{\bf A691}  (2001) 
779,[arXiv:hep-ph/0012167];\,
{\bf B573} (2000) 833,  [arXiv:hep-ph/9908317].


\bibitem{EKL}
R.~K.~Ellis, Z.~Kunszt and E.~M.~Levin, {\it
Nucl.\ Phys.}  {\bf B420} (1994) 517
[Erratum-ibid.\  {\bf B433} (1995) 498].

\bibitem{GLMSOLB}
E.~Gotsman, M.~Kozlov, E.~Levin, U.~Maor and E.~Naftali,
{\it ``Towards a new global QCD analysis: Solution to the non-linear 
equation 
at
arbitrary impact parameter,''}
arXiv:hep-ph/0401021.
\bibitem{LUB}
M.~Lublinsky,
{\it Eur.\ Phys.\ J.}\,   {\bf C21} (2001) 513
[arXiv:hep-ph/0106112].

\bibitem{GBSSOL}
K.~Golec-Biernat and A.~M.~Stasto, {\it 
Nucl.\ Phys.}\, {\bf B668} (2003) 345
[arXiv:hep-ph/0306279].


\bibitem{IKMC}
Takashi Ikeda and Larry McLerran, {\it ``Impact parameter dependence in
the
Balitsky-Kovchegov equation"},\,\,
arXiv:hep-ph/0410345.

\bibitem{ALAR}
J.~L.~Albacete, N.~Armesto, J.~G.~Milhano, C.~A.~Salgado and
U.~A.~Wiedemann,
arXiv:hep-ph/0408216.


\end{thebibliography}
\end{document}